\documentclass[aps,preprint,tightenlines,superscriptaddress,amsmath,amssymb]{revtex4-2}
\usepackage[T1]{fontenc}
\usepackage[utf8]{inputenc}
\usepackage{graphicx}
\usepackage{hyperref}
\usepackage{float}
\usepackage{xcolor}
\usepackage{siunitx}
\ifdefined\qty\else
  \ifdefined\NewCommandCopy
    \NewCommandCopy\qty\SI
  \else
    \NewDocumentCommand\qty{O{}mm}{\SI[#1]{#2}{#3}}
  \fi
\fi
\ifdefined\qtyproduct\else
  \ifdefined\NewCommandCopy
    \NewCommandCopy\qtyproduct\SI
  \else
    \NewDocumentCommand\qtyproduct{O{}mm}{\SI[#1]{#2}{#3}}
  \fi
\fi
\ifdefined\qtyrange\else
  \ifdefined\NewCommandCopy
    \NewCommandCopy\qtyrange\SI
  \else
    \NewDocumentCommand\qtyrange{O{}mmm}{\SIrange[#1]{#2}{#3}{#4}}
  \fi
\fi

\begin{document}

\title{Grating-based microcavity with independent control of resonance energy and linewidth for non-Hermitian polariton system}

\author{Jiaqi Hu}
  \affiliation{Applied Physics Program, University of Michigan, Ann Arbor, Michigan 48109, USA}
\author{Nathanial Lydick}
\author{Zhaorong Wang}
  \affiliation{Department of Physics, University of Michigan, Ann Arbor, Michigan 48109, USA}
\author{F. Jabeen}
  \affiliation{Technische Physik, Physikalisches Institut and W\"urzburg-Dresden Cluster of Excellence ct.qmat, Universit\"at W\"urzburg, 97074 W\"urzburg, Germany}
\author{C. Schneider}
  \affiliation{Technische Physik, Physikalisches Institut and W\"urzburg-Dresden Cluster of Excellence ct.qmat, Universit\"at W\"urzburg, 97074 W\"urzburg, Germany}
  \affiliation{Institute of Physics, University of Oldenburg, 26129 Oldenburg, Germany}
\author{S. H\"ofling}
  \affiliation{Technische Physik, Physikalisches Institut and W\"urzburg-Dresden Cluster of Excellence ct.qmat, Universit\"at W\"urzburg, 97074 W\"urzburg, Germany}
\author{Hui Deng}
  \email{dengh@umich.edu}
  \affiliation{Applied Physics Program, University of Michigan, Ann Arbor, Michigan 48109, USA}
  \affiliation{Department of Physics, University of Michigan, Ann Arbor, Michigan 48109, USA}

\date{\today}

\begin{abstract}
Exciton-polaritons have become an emerging platform for implementing non-Hermitian physics. The implementation commonly requires control of both the real and imaginary parts of the eigenmodes of the system. 
We present an experimental method to achieve this purpose using microcavities with sub-wavelength gratings as reflectors. 
The reflectivity and reflection phase of the grating can be changed by its geometric parameters and they determine the energy and linewidth of the polariton modes. 
We demonstrate that this method allows a wide range of possible polariton energy and linewidth, suitable for implementing non-Hermitian polariton systems with coupled modes. 
\end{abstract}

\maketitle

Non-Hermitian physics \cite{bender_making_2007} has been recently studied in many optical systems, which naturally involve gain and loss \cite{longhi_parity-time_2017, el-ganainy_non-hermitian_2018}. Novel phenomena are expected near exceptional points, which are singularities of the spectrum in the parameter space spanned by the complex energies of different modes within the system and the coupling between the modes. Experimental observations range from uni-directional propagation \cite{lin_unidirectional_2011, feng_experimental_2013} to orbital angular momentum beam emission \cite{miao_orbital_2016} and enhanced sensing \cite{chen_exceptional_2017}. 

A particularly interesting system for non-Hermitian physics is the exciton-polariton system, which allows the study of the interplay of non-Hermitian physics with nonlinearity-induced features such as superfluidity
\cite{deng_exciton-polariton_2010,fraser_coherent_2017, ghosh_quantum_2020}. Exciton-polaritons are formed by strong-coupling between a photonic cavity mode and an exciton mode, each subject to gain and loss.
Compared to typical optical systems, the exciton component introduces a much stronger nonlinearity.  
In a uniform polariton system, spontaneous condensation into different modes has been predicted depending on the phase space trajectory around the exceptional points \cite{hanai_non-hermitian_2019}, and adiabatic flipping between the photonic and excitonic modes was envisioned near the exception points \cite{khurgin_exceptional_2020}. These studies provide new insight to unique features of non-equilibrium phase transitions.  
Coupling multiple laterally localized polariton modes provides further freedom to engineer the non-Hermitian system and to access novel phenomena such as non-Hermitian skin effect due to effective non-reciprocal hopping \cite{mandal_nonreciprocal_2020, xu_interaction-induced_2021, xu_nonreciprocal_2021} and synchronization among a large array via supersymmetry selection \cite{hokmabadi_supersymmetric_2019}. 

Exploring these phenomena often requires precise tuning of the system parameters, including the real and imaginary part of energies of the individual modes and the coupling strength between them. 
Such tuning is difficult with conventional cavities made of distributed Bragg reflectors (DBRs). While the cavity energy is readily tunable by a tapered cavity thickness, other parameters, such as the cavity decay rate, exciton energy and decay rate, and exciton-photon coupling, are generally not tunable in the same sample. 
Creating a polariton lattice, by etched micropillars or patterned metal layer deposited on top of the cavity \cite{schneider_exciton-polariton_2017}, introduces a variable coupling among the localized polariton modes, yet it is challenging to decouple the energy and linewidth of the modes. 
Alternatively, spatially structured non-resonant optical pumps have also been used to introduce a potential landscape and provide polariton gain, leading to observations such as topological Berry phase \cite{gao_observation_2015}, chiral state emission \cite{gao_chiral_2018}, condensation in topological defect state \cite{pickup_synthetic_2020}, and optical switching of condensates \cite{li_switching_2021}. However, the pump introduces non-uniform and nonlinear modulations of multiple parameters of the system simultaneously. Understanding of the system properties often requires comparison between experiments and simulation for each device and pump configuration, making it difficult for \textit{a priori} predictions as well as extensions to more complicated or large-scale lattices.  

Here we demonstrate a method to independently control both the energy and linewidth of the photonic cavity mode, and consequently the polariton mode. 
It is based on a hybrid microcavity with a sub-wavelength grating and a DBR as the two reflectors \cite{zhang_zero-dimensional_2014}. 
The reflectivity and reflection phase of the grating, which affect the cavity decay rate and resonant energy respectively, can be tuned via the geometric parameters of the grating. 
The complex eigenenergy of the polariton is determined in a completely passive way during the fabrication of the device, which allows the device to be compatible with simple spatially uniform optical pumping or even electrical pumping. 
The gratings also define the in-plane spatial extent of the polariton mode, therefore allowing the creation of multiple localized modes with controllable coupling strengths that could be potentially extended to large scale devices, as demonstrated in Refs.~\cite{zhang_coupling_2015, kim_emergence_2020}. 
The polariton mode will be linearly polarized for the gratings we are using and the polarization is determined by the orientation of the grating. 
We demonstrate devices with a tuning range on the order of a few meV for the eigenenergy and from sub-meV to a few meV for the linewidth, appropriate for models that require the energy, linewidth, and coupling strength to be comparable \cite{hanai_non-hermitian_2019, khurgin_exceptional_2020, comaron_non-hermitian_2020}.

\begin{figure}[h]
\begin{center}
\includegraphics[width=0.5\textwidth]{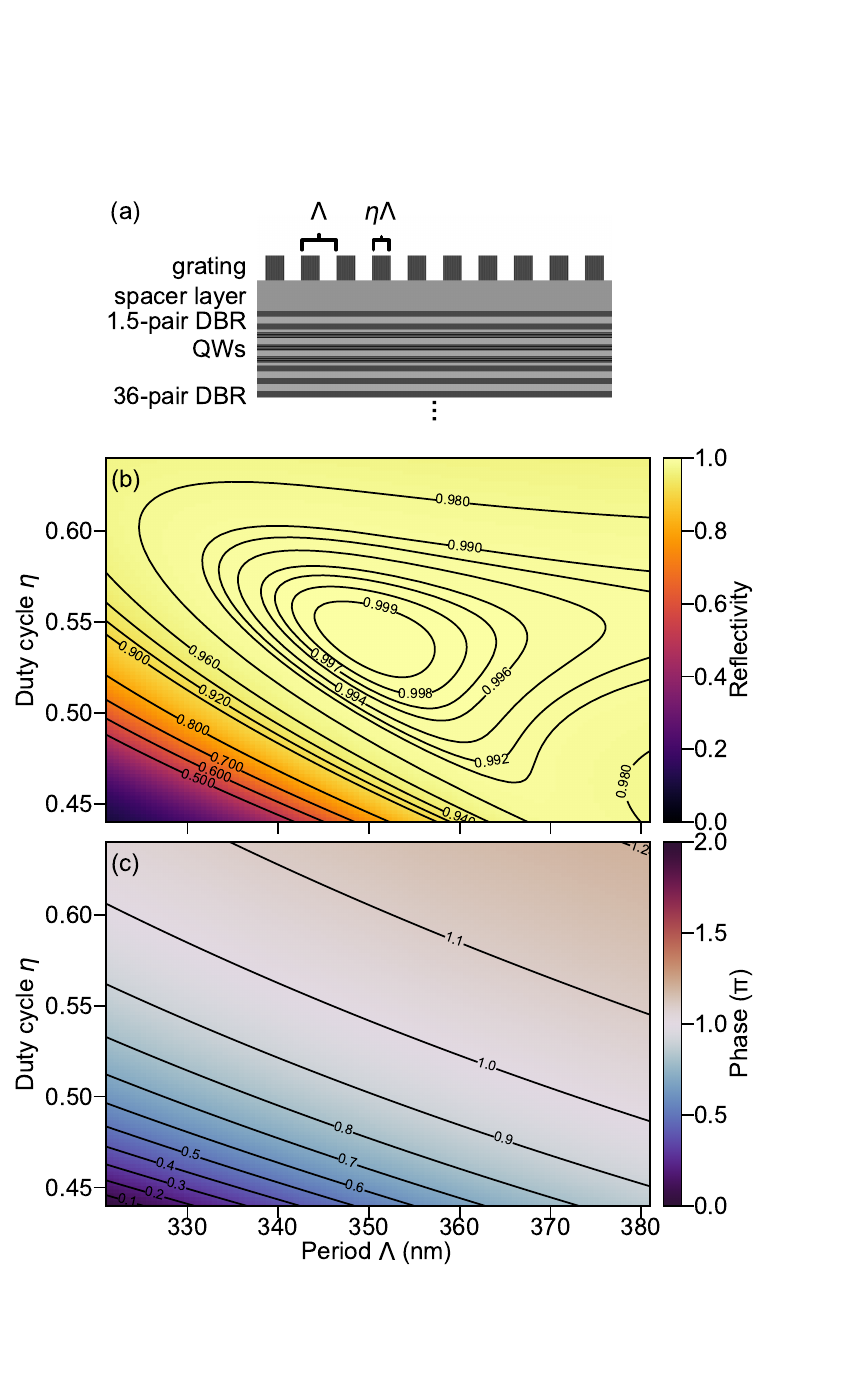}
\caption{
(a) A schematic of the cross-sectional profile of the microcavity. 
Simulated zeroth-order reflectivity (b) and reflection phase (c) of a grating-based mirror as a function of the period and duty cycle of the grating, with \qty{800}{nm} light polarized perpendicularly to the grating bars and normal incidence from the $\mathrm{AlAs}$ side. 
}
\label{fig:grating_rp}
\end{center}
\end{figure}

High-contrast sub-wavelength gratings have been used as broad-band high reflectivity mirrors in cavities of both vertically emitting lasers and exciton polaritons \cite{qiao_recent_2018, zhang_zero-dimensional_2014}. 
Particularly, for a square binary grating, whose cross-sectional profile is shown in Fig.~\ref{fig:grating_rp}(a), the mechanism of the high reflectivity can be understood as the cancelation of the zeroth-order Fourier component of two waveguide-array-like resonant modes of the grating structure at the output (transmission) interface \cite{karagodsky_theoretical_2010}. 
The resonant frequency and field amplitude distribution of each of the modes are sensitive to the geometry of the grating bars with different dependencies. 
Therefore, a change in the period or duty cycle of the grating results in an overall change of the amplitude and phase of the reflected field. This allows us to tune the energy and linewidth of the cavity resonance locally.

To evaluate the accessible range of tuning, we first perform numerical simulations 
\footnote{We use the numerical solver from Ref.~\cite{liu_s4_2012}.} 
with the Rigorous Coupled-Wave Analysis (RCWA) method \cite{moharam_rigorous_1981} for a realistic grating mirror structure. 
The mirror consists of an $\mathrm{Al_{0.2}Ga_{0.8}As}$ grating on top of an $\mathrm{Al_{0.85}Ga_{0.15}As}$ layer, followed by a 1.5-pair DBR, with the DBR side facing an $\mathrm{AlAs}$ cavity. 
Fig.~\ref{fig:grating_rp}(b, c) shows an example of the zeroth-order reflectivity and reflection phase near a high-reflectivity region. 
The reflection is computed for \qty{800}{nm} light with the electric field polarized perpendicularly to the grating bars and normal incidence from the $\mathrm{AlAs}$ side. 
For the one-dimensional gratings, the reflection properties can typically only be optimized for one of the orthogonal linear polarizations but not both simultaneously. 
In this example, for the polarization parallel to the grating bars the reflectivity in this parameter range is only about \num{0.5}. 
The grating has a thickness of \qty{242}{nm} and is optimized for high reflectivity at \qty{800}{nm}. 
The period $\Lambda$ and duty cycle $\eta$ are the geometric tuning parameters and the highest reflectivity is obtained at $(\Lambda_0, \eta_0) = (\qty{351}{nm}, \num{0.54})$. 
The thickness of the $\mathrm{Al_{0.0.85}Ga_{0.15}As}$ layer is chosen so that the reflection phase is $\pi$ at the highest reflectivity. 

As $(\Lambda, \eta)$ are tuned away from $(\Lambda_0, \eta_0)$ in the two-variable parameter space, the reflectivity decreases [Fig.~\ref{fig:grating_rp}(b)] while the reflection phase [Fig.~\ref{fig:grating_rp}(c)] changes monotonously in the vicinity.
Therefore, by selecting certain combinations of period and duty cycle, it is possible to attain reflectivity and phase separately. 
The available ranges of the reflectivity and phase are not independent. 
With Fig.~\ref{fig:grating_rp}(b, c) as an example, given a fixed reflectivity, the corresponding $(\Lambda, \eta)$ can only move on a contour line, corresponding to a limited range of phase. 
However, as we will show later, the ranges are often adequate. 

In experiment, the thickness of the layers of the wafer as well as the parameters of the grating may differ from the design, which can be calibrated through iterations between simulation and measurement of the energy and linewidth.

We fabricate the grating-based cavity devices on $\mathrm{GaAs}$-based epitaxial wafers. 
These wafers have 12 $\mathrm{GaAs/AlAs}$ quantum wells, each of thickness \qty{12}{nm}, distributed at three center anti-nodes of the resonant mode of an $\mathrm{AlAs}$ $3\lambda/2$-cavity. 
The lower mirror of the cavity is a $\mathrm{Al_{0.15}Ga_{0.85}As/AlAs}$ or $\mathrm{Al_{0.2}Ga_{0.8}As/AlAs}$ DBR with a total of 36 pairs of layers. 
The upper mirror is the aforementioned assembly of $\mathrm{Al_{0.2}Ga_{0.8}As}$ grating layer, an $\mathrm{Al_{0.85}Ga_{0.15}As}$ spacer layer, and 1.5 pairs of DBR. 
Each grating device is in a \qtyproduct{25x25}{\um} or \qtyproduct{30x30}{\um} square region, and many devices are made with different periods and duty cycles near the reflectivity maximum. 
The pattern is first written with e-beam lithography on a spin-coated ZEP520A e-beam resist layer of about \qty{300}{nm} thickness, and then transferred to the grating layer with $\mathrm{BCl_3}$ reactive ion etching. 

\begin{figure}[h]
\begin{center}
\includegraphics[width=0.5\textwidth]{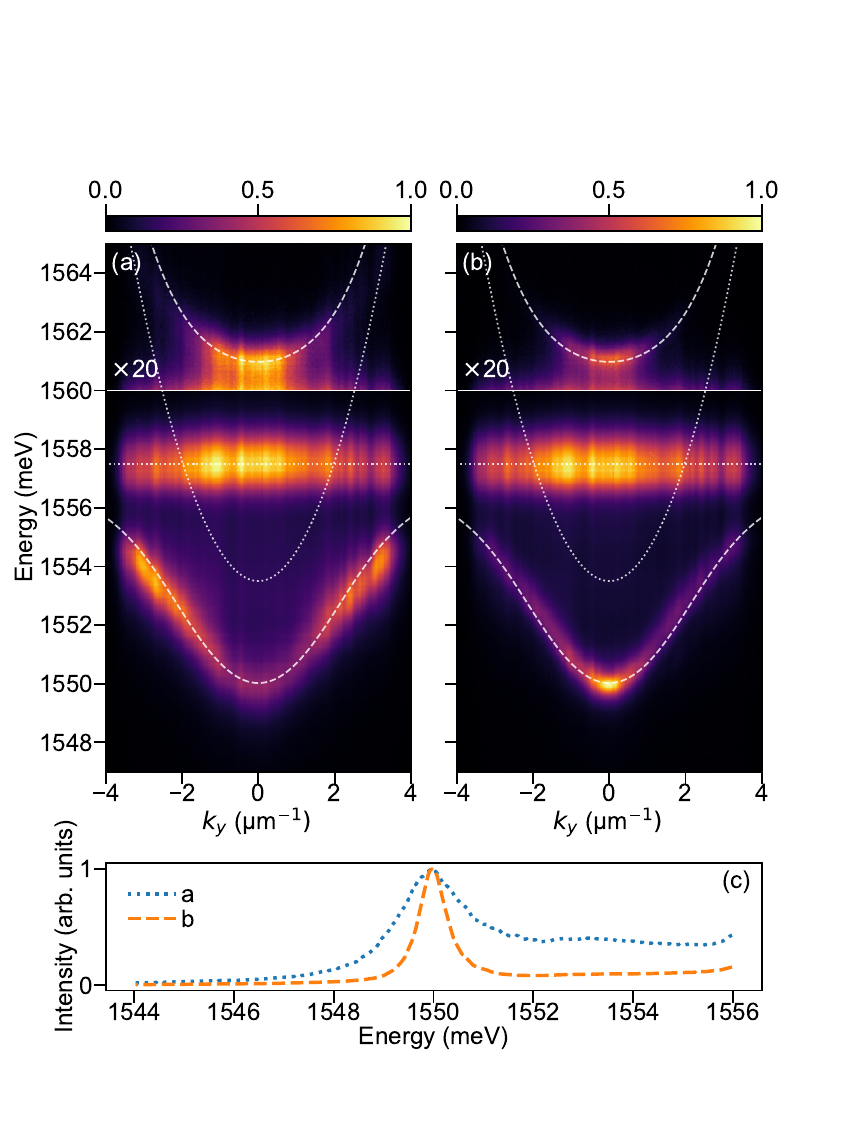}
\caption{
Fourier space PL spectra of typical devices showing equal lower polariton energy but different linewidth. 
(a) Grating with designed period \qty{324}{nm} and duty cycle \num{0.763}. 
(b) Grating with designed period \qty{348}{nm} and duty cycle \num{0.763}. 
In subfigures (a) and (b), the horizontal axes are the wave vector parallel to the grating bars ($k_y$). 
The spectra are collected with a polarizer perpendicular to the grating bars. 
The dashed lines are the polariton dispersions of a conventional DBR-DBR microcavity for comparison, calculated with a coupled oscillator model, assuming a quadratic cavity resonance dispersion.
The parameters in the calculation are chosen to obtain the same exciton and polariton energies at zero wave vector as the measured devices, and the uncoupled exciton and cavity energies are marked with the dotted lines. 
(c) Line spectra near the lower polariton energy corresponding to (a) and (b), integrated within $k_y < \qty{0.1}{\um^{-1}}$ and normalized to the maximum value in the energy range. 
}
\label{fig:spectra}
\end{center}
\end{figure}

Fig.~\ref{fig:spectra} shows photoluminescence (PL) spectra of typical devices in Fourier space, with a polarizer perpendicular to the grating bars. 
The devices are pumped with a non-resonant continuous-wave laser focused to a \qty{2}{\um} spot in the center of the grating. 
Strong-coupling is evidenced by the emission from lower polariton and upper polariton branches for both devices in display. 
There is emission from near the exciton energy, due to the inhomogeneous broadening of the exciton \cite{houdre_vacuum-field_1996} or emission from excitons with large wave vectors through first order diffraction of the grating. 
For the orthogonal polarization, the excitons remain in the weak-coupling regime. 

These two devices provide an example of local linewidth control by the grating parameters. Their design parameters are $\Lambda = \qty{324}{nm}$, $\eta = \num{0.763}$ [Fig.~\ref{fig:spectra}(a)] and $\Lambda = \qty{348}{nm}$, $\eta = \num{0.763}$ [Fig.~\ref{fig:spectra}(b)]. The resulting lower polariton modes have nearly the same energy at zero wave vector at \qty{1549.98\pm0.02}{meV} and \qty{1549.99\pm0.01}{meV}, but the linewidths are vastly different at \qty{1.67\pm0.07}{meV} and \qty{0.67\pm0.01}{meV} [Fig.~\ref{fig:spectra}(c)]. 
The Rabi splitting is $\hbar\Omega_R \approx \qty{10}{meV}$ at zero wave vector, computed from the exciton energy measured in the unetched area near the device and both the upper and lower polariton energies. 
The dispersions of the cavity, and thus of the polaritons, have a small deviation from that of a conventional DBR-DBR microcavity with quadratic cavity resonance dispersion because the grating reflection phase also depends on the wave vector. 
The polariton dispersions of a conventional DBR-DBR microcavity are calculated with a coupled oscillator model and overlaid on the spectra for comparison, with parameters that result in the same exciton and polariton energies at zero wave vector as the measured devices and assuming a quadratic cavity resonance dispersion.

\begin{figure}[h]
\begin{center}
\includegraphics[width=0.5\textwidth]{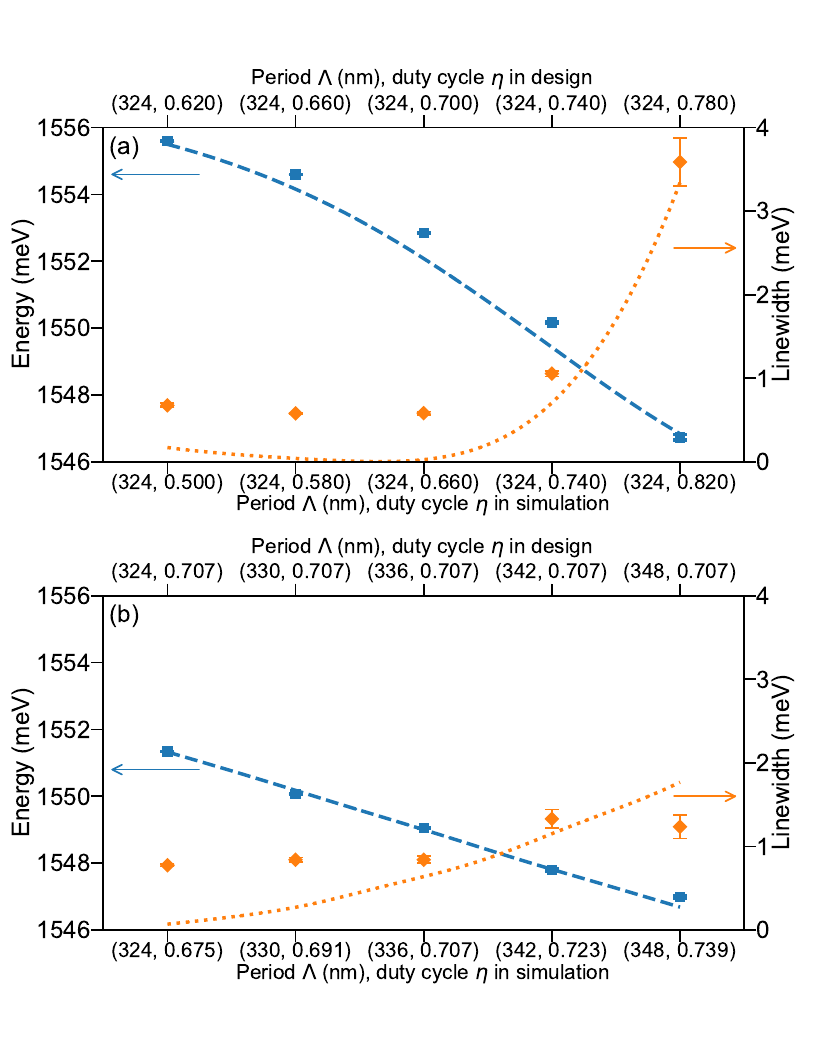}
\caption{
Measured and simulated lower polariton energy and linewidth at zero wave vector, for two series of gratings shown in (a) and (b) respectively. 
(a) shows gratings designed to have the same period and varying duty cycles, while (b) shows varying period and same duty cycle. 
The upper horizontal axis labels the designed parameters while the lower horizontal axis labels the parameters used in simulation. 
The experimental energy (blue square markers) and linewidth (orange diamond markers) values are obtained by curve fitting of the PL spectra with Lorentzian peaks and the error bars represent the uncertainty of the corresponding fitting parameters calculated from the diagonal elements of the covariance matrix of the parameters. 
The simulated energy (dashed lines) and linewidth (dotted lines) values are obtained by curve fitting of the simulated reflection spectra with Fano lineshape. 
}
\label{fig:step}
\end{center}
\end{figure}

To evaluate experimentally viable step sizes of the individual parameters, we fabricate arrays of gratings with a range of different periods $\Lambda$ and duty cycles $\eta$; we then compare the design parameter values with values measured through scanning electron microscope (SEM), and compare the polariton energy and linewidth of these devices in RCWA simulated reflection spectra with measured values in PL [Fig.~\ref{fig:step}]. 
All RCWA simulations are performed with $41$ basis functions and numerical convergence is verified. 

The period $\Lambda$ of the fabricated devices follows the design very well apart from a \qty{2.8\pm0.4}{\percent} offset above the designed values.
This is consistent among devices with different parameters and is likely from an overall error in the calibration of the magnification of either the lithography or the measurement. 
The measured duty cycle $\eta$ sometimes has a larger deviation from the design, due to the limitation of beam positioning accuracy of the e-beam lithography and due to insufficient anisotropy the etching process. 
For example, when the duty cycle is kept the same in the design and only the period is changed (near \qty{340}{\nm}), the duty cycle can still vary by \num{0.07} (about \qty{24}{\nm}). 
If the duty cycle is also varied in design, the discrepancy can be even larger, though the measurement uncertainty of the duty cycle is limited by the SEM to about \qty{20}{nm}. 

To calibrate the actual duty cycle more accurately, we compare the measured energy and linewidth of the devices with RCWA simulations where we fix the period at the designed value and allow duty cycle to be scaled linearly. 
As shown in Fig.~\ref{fig:step}, the measured linewidth follows qualitative the simulated results but are generally larger than the simulated linewidth. This is expected as the linewidth of the device is susceptible to imperfections in fabrication, such as sidewall roughness and surface scattering. 
The finite size of the gratings can also contribute to the broadening although it is not expected to be dominant here because the size of the devices is larger than $50$ periods of grating \cite{zhou_large_2008, tibaldi_high-contrast_2015}. 
On the other hand, the polariton energies show excellent agreement between simulation and measurements. 

From this result we estimate a step size of \qty{6}{nm} for the period and \num{0.08} (\qty{26}{\nm}) for the duty cycle. For our devices, these step sizes translate to polariton energy resolution of about \qty{1}{meV} and linewidth resolution down to \qty{0.5}{meV}. 
By mixing changes in both parameters, it is possible to obtain finer steps of polariton energy or linewidth. 
We also expect that the duty cycle step size can be further reduced to $\lesssim\qty{6}{nm}$ with improvements to the fabrication process.

\begin{figure}[h]
\begin{center}
\includegraphics[width=0.5\textwidth]{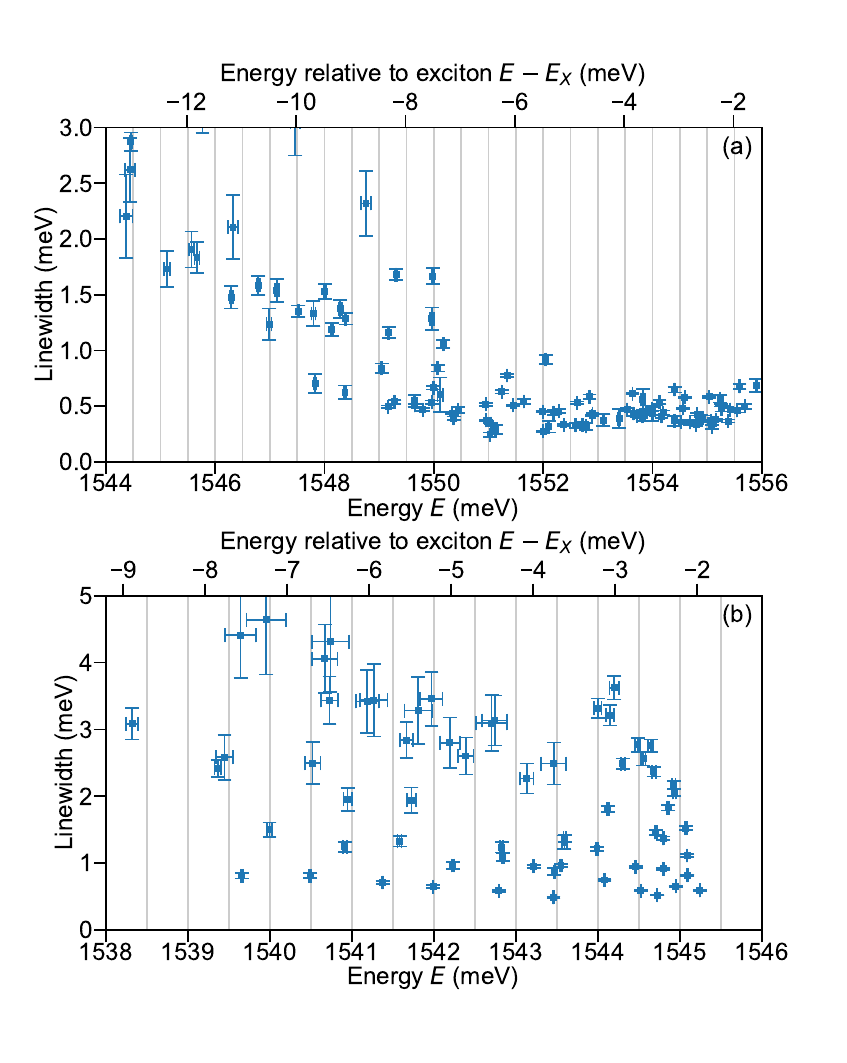}
\caption{
A collection of measured PL energy and linewidth of lower polariton modes at zero wave vector of gratings with various parameters on two different sample pieces. 
(a) The gratings have designed periods around \qty{336}{nm} and duty cycles around \num{0.72}. 
(b) The gratings have designed periods around \qty{550}{nm} and duty cycles around \num{0.50}. 
Gray vertical lines mark \qty{0.5}{meV} energy spacing. 
The experimental energy and linewidth values are obtained by curve fitting of the spectra with Lorentzian peaks and the error bars represent the uncertainty of the corresponding fitting parameters calculated from the diagonal elements of the covariance matrix of the parameters. 
}
\label{fig:ew}
\end{center}
\end{figure}

To evaluate how broadly we can separately tune the energy and linewidth, we fabricate arrays of gratings of varying $\Lambda$ and $\eta$ on \qtyproduct{5x5}{mm} pieces of wafers. Fig.~\ref{fig:ew}(a) and (b) show the measured PL energy and linewidth of lower polariton modes at zero wave vector from the gratings on two examples pieces respectively.

In general, the ranges of energy and linewidth are limited by the choice of the other value. 
To obtain a certain detuning, for example, will require the parameters to stay on a curve in the parameter space that gives the grating mirror the correct reflection phase for that cavity energy, and the reflectivity can only take values on that curve, limiting the linewidth. 
On the other hand, the reflectivity needs to be maintained at high enough values for strong-coupling and hence limits the range of detuning. 
For the piece shown in Fig.~\ref{fig:ew}(a), the energy range spans more than \qty{10}{meV}. 
For energies below \qty{1550}{meV}, the linewidth range spans over \qty{1.5}{meV}. 
For energies above \qty{1550}{meV}, the linewidth is narrower and ranges between \qty{0.24}{meV} and \qty{0.92}{meV}. 
The reduced linewidth range is due to more positive detuning and smaller photonic fraction at these energies, as the linewidth variation comes from the photonic mode. 
Results from a similar wafer piece are shown in Fig.~\ref{fig:ew}(b), where high-reflectivity is found in the parameter region near $\Lambda\sim\qty{550}{nm}$ and $\eta\sim\num{0.5}$. 
In this piece, for more than \qty{5}{meV} of energy range, the linewidth can be tuned by more than \qty{2}{meV}. 

With these records as a calibration, a practical procedure can be followed to design and fabricate new devices that require local modes with different complex eigenenergies: first determine the target energy and linewidth, and then select the corresponding parameters that achieve those values from the present result.

Finally, with the tuning capabilities demonstrated, we consider the possibility of realizing a simple design of the non-Hermitian device that consists of two coupled local polariton modes. 
For such devices, it is often required that the coupling strength and the difference between the imaginary part of energies are comparable. 
It has been shown that sub-wavelength grating-based cavities can confine polariton modes in-plane to smaller than \qty{10}{\um}, and that these modes can be coupled to form dimer-like modes or one-dimentional chains \cite{zhang_zero-dimensional_2014, zhang_coupling_2015}. 
The coupling strength between neighboring modes is on the order of \qty{0.5}{meV} \cite{kim_emergence_2020}, comparable to the tuning range of linewidth in the present case. 
In Fig.~\ref{fig:ew} vertical lines with \qty{0.5}{meV} spacing are added in the background to show the available choices of linewidth within \qty{0.5}{meV} energy bins.

In summary, we have demonstrated that grating-based microcavities can be used to create polariton modes with controllable energy and linewidth. The polariton energy can be tuned over a range comparable to that of the normal mode splitting, while the linewidth can be independently tuned from a fraction of a meV to a few meV. The tuning resolution of \qtyrange{0.5}{1}{meV} is mainly limited by the e-beam lithography resolution of \qty{24}{nm}. Improvements in the fabrication accuracy, especially in the e-beam lithography, will enable finer tuning resolution. 
The demonstrated system is scalable to large lattices. 
The ability to accurately and independently tune the energy, loss and gain of the system may facilitate the study a wide range of non-Hermitian phenomena on the polariton platform \cite{kawabata_symmetry_2019, hanai_non-hermitian_2019, khurgin_exceptional_2020, comaron_non-hermitian_2020}, and may also enable development of single photon or entangled photon sources based on coupled, driven polariton systems \cite{liew_single_2010, bamba_origin_2011, liew_multimode_2013}.

\begin{acknowledgments}
J.H., N.L., Z.W., and H.D. acknowledge financial support from the US National Science Foundation (NSF) under Grant No. DMR 2004287. 
C.S. acknowledges funding by the German Research Foundation (DFG)
within the Project No. SCHN1376/13-1 and S.H. within the Project No. HO5194/12-1. 
This work was performed in part at the University of Michigan Lurie Nanofabrication Facility. 
We thank S. Brodbeck and P. Wyborski for their contributions to this work by epitaxial growth of samples. 
J.F., C.S., and S.H. gratefully acknowledge support of this work by the Free State of Bavaria. 
\end{acknowledgments}

\section*{Conflict of Interest}

The authors have no conflicts to disclose.

\section*{Author Contributions}

{\bf Jiaqi Hu:} Conceptualization (equal); data curation (equal); formal analysis (equal); investigation (lead); methodology (equal); validation (equal); visualization (equal); writing -- original draft (lead); writing -- review and editing (equal). 
{\bf Nathanial Lydick:} Investigation (supporting); writing -- review and editing (equal). 
{\bf Zhaorong Wang:} Resources (equal). 
{\bf F. Jabeen:} Resources (equal). 
{\bf C. Schneider:} Resources (equal); writing -- review and editing (equal). 
{\bf S. H\"ofling:} Resources (equal); writing -- review and editing (equal). 
{\bf Hui Deng:} Conceptualization (equal); funding acquisition (equal); project administration (equal); supervision (equal); writing -- original draft (supporting); writing -- review and editing (equal).

\section*{Data Availability Statement}

The data that support the findings of this study are available from the corresponding author upon reasonable request. 

\bibliography{grating_tuning}

\end{document}